\documentclass[aps,preprint,groupedaddress,showpacs]{revtex4-1}

\usepackage[labelfont=bf,labelsep=period,justification=raggedright]{caption}
\usepackage{graphicx}
\usepackage{color}
\usepackage{comment}
\usepackage{subfig}
\usepackage{amsmath}

\begin{document}


\title{Sleeping Beauties in Meme Diffusion}


\author{Leihan Zhang\textsuperscript{1}, Ke Xu\textsuperscript{1} and Jichang Zhao\textsuperscript{2,*}}
\affiliation{$1$ State Key Lab of Software Development Environment, Beihang University\\
$^1$ School of Economics and Management, Beihang University \\
$^\star$Corresponding author: jichang@buaa.edu.cn}


\date{\today}

\begin{abstract}
A sleeping beauty in diffusion indicates that the information, can be ideas or innovations, will experience a hibernation before a sudden spike of popularity and it is widely found in citation history of scientific publications. However, in this study, we demonstrate that the sleeping beauty is an interesting and unexceptional phenomenon in information diffusion and even more inspiring, there exist two consecutive sleeping beauties in the entire lifetime of propagation, suggesting that the information, including scientific topics, search queries or Wikipedia entries, which we call memes, will go unnoticed for a period and suddenly attracts some attention, and then it falls asleep again and later wakes up with another unexpected popularity peak. Further explorations on this phenomenon show that intervals between two wake ups follow an exponential distribution and the second wake up generally reaches its peak at a higher velocity. In addition, higher volume of the first wake up will lead to even much higher popularity of the second wake up with great odds. Taking these findings into consideration, an upgraded Bass model is presented to well describe the diffusion dynamics of memes on different media. Our results can help understand the common mechanism behind propagation of different memes and are instructive to locate the tipping point in marketing or find innovative publications in science.

\end{abstract}


\maketitle


\section{Introduction}
\label{sec:intro}

Meme is usually defined as the simplest cultural unit that spreads between different individuals and may gain collective attention within a community or culture~\cite{Dawkins1976,Christian2011}. Dawkins even postulates meme as a cultural analogy of genes in order to explain how innovations, ideas, catchphrases, melodies, rumors, or fashion trends disseminate through a population~\cite{Dawkins1976}. In recent decades, Internet and its various applications, like pubMeds, Wikipedia and online social media, provide massive digital fossils of meme diffusion, which offer us a decent proxy to disclose the mechanism beyond the propagation. The insights from these investigations can help us understand the rules that produce the dynamics and establish models that estimate the dynamics of trends.

Basic dynamics of the meme diffusion within the same media has been comprehensively studied from different perspectives. For example, mathematical epidemiology as well as simple log-normal distributions are suggested to profile the growth and decline of diffusion~\cite{Leskovec2009,Christian2011,Sasahara2013,Spitzberg2014}, how competition, homogeneity and network cooperatively affect the spread is discussed~\cite{Gleeson2014,Gleeson2015,Carlos2015,Centola1194,Centola1269,cheng2014can}, the different roles in diffusion played by different individuals are revealed by~\cite{Bentley2012} and~\cite{Zhang2016} and even simulation models are established to replicate the meme diffusion in Twitter~\cite{weng2012,weng2013,weng2014,bauckhage2013mathematical,Wang2015,Yu2016}. However, except disclosing the common features of successful memes in different online social networks~\cite{Shifman2012,Michele2013,Coscia2014,Ceyda2015}, the universal mechanism that essentially drives the propagation of memes in different media still remains unclear. In this study, we argue that the sleeping beauties existing in the lifetime of different memes can be a path to unravel the common knowledge behind diffusions of different media. 

Sleeping beauty, exhibiting a hibernation before an unexpected popularity peak, is pervasively found and studied in the diffusion of memes like ideas or innovations in scientific publications. Garfield first provide examples of articles with delayed recognition~\cite{Garfield1980,Garfield1989}, which can be identified through the citation history~\cite{Van2004,better2005}. ~\cite{Van2004} then coined the term ``sleeping beauty'' in reference to the delayed recognition and several basic features, including length of sleep, depth of sleep and awake intensity are proposed to measure sleeping beauties. Later, finding general features of sleeping beauties~\cite{Redner2005,Ohba2012,Li2012sciento,Marx2014} and explaining the awakening reasons in paper citations~\cite{Braun2010,Bornmann2014,Li2014,Lachance2014,Burrell2015} attract most of the attention. Indeed, understanding the sleeping beauty in science will help improve the impact factor and mine the surprising innovation~\cite{van2015pone}, however, most of the previous studies only focus on the scientific publications and ignore the possibility that in other memes, like trending topics in social media, hot queries in Google or popular entries in Wikipedia, might also experience the similar sleeping beauty, which in fact greatly motivates the present work. 

Actually the existing evidence already implicitly implies the latent connection between different memes in terms of sleeping beauties. For scientific papers, ~\cite{Li2013citation} find that there are some papers appearing ``flash in the pan'' first and then ``delay recognition'', i.e., these papers experience two sleeping beauties in their citation history. More uplifting is that the Internet slang words also demonstrate the same phenomenon in Weibo~\cite{Zhang2016}. This similarity suggests that for a meme, no matter it is an idea in scientific citations or a slang word in social media, its diffusion might be subject to a common rule. In fact, many long-lived memes in different media for example, a name, an event or an idea, experience sleeping beauties more than twice. Therefore, comparison of different sleeping beauties for one same meme will be a decent perspective to probe into the common rule behind meme propagation. And further we believe that the prior sleeping beauty can help predict the same meme's next wake up. And considering that the higher popularity usually means wider diffusion and greater influence, then we will focus on the two remarkable sleeping beauties of memes having the characteristics of sleeping beauties, one cascade with global highest popularity and the other cascade with local highest popularity in front of the highest cascade. With the basic assumption that different memes in different media share the same diffusion mechanism, we try to disclose the possible common rule through the in-depth investigation on the phenomenon of sleeping beauties. And the knowledge of this common rule can help upgrade the existing simulation or prediction approaches and extend them to many different domains.

\begin{figure}[ht]
\centering
\includegraphics[width=\linewidth]{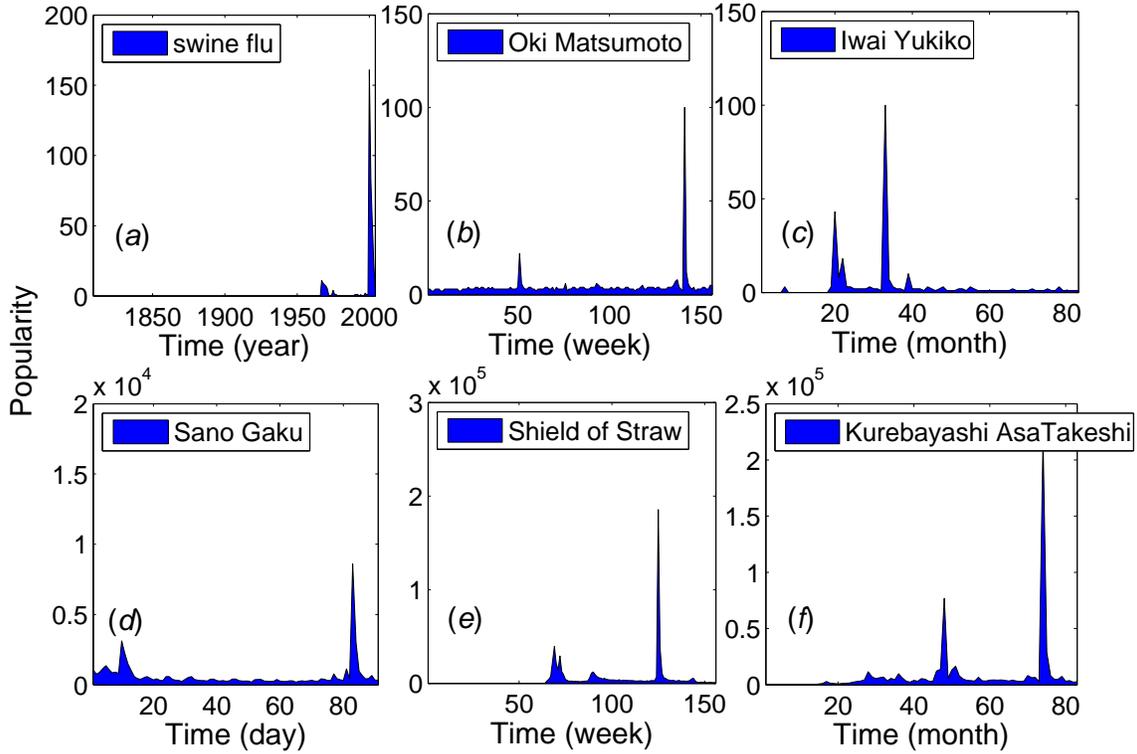}
\caption{ The popularity dynamics of six memes from different datasets. For convenience, we use popularity to denote the utilization volume, which is generally considered as the degree of public concern on a particular meme. (\emph{a}) The yearly popularity of 2-grams ``swine flu'' from 1809 to 2013. (\emph{b} and \emph{c}) The relative search volume of two queries, ``Oki Matsumoto'' and ``Iwai Yukiko'' in Google and the time granularities are respectively week and month. (\emph{d}, \emph{e} and \emph{f}) The Wikipedia page views of three items, including ``Sano Gaku'', ``Shield of Straw'' and ``Kurebayashi AsaTakeshi'', and the multiple time granularities are respectively set to day, week, and month.}
\label{fig1}
\end{figure}

In order to systematically explore the sleeping beauties in different memes' diffusion, we investigate three datasets, including utilization volume of $n$-grams in scientific publication titles during a long period, search queries in Google, and page view statistics of Wikipedia entries. That's to say, the memes we focus on come from different backgrounds, which guarantees the generalization of our following findings. Note that the time granularities in meme diffusion are also diverse, including year, month, week and day for different datasets and it further ensures that our study can discuss the entire lifetime of the propagation from a long-term perspective. As shown in Fig.~\ref{fig1}, six typical memes are sampled to demonstrate their diffusion dynamics with different time granularities. As can be seen, these memes experience two obvious sleeping beauties, which means that each of them goes unnoticed for a period and suddenly attracts considerable attention, and then falls asleep again, following with another unexpected popularity spike higher than the previous one. This interesting phenomenon of two sleeping beauties is independent of the media and time granularities, which has not been discussed in the previous work to our best knowledge. And it raises many unsolved but fundamental questions like how to identify these two beauties, how they distribute in lifetime of the diffusion and can they be used to predict the future trends.

Inspired by the concept of beauty coefficient proposed by ~\cite{Ke2015}, we introduce a framework to identify two sleeping beauties in each meme's diffusion. We demonstrate that the phenomenon of two sleeping beauties is pervasively existing in diffusion of different memes and time intervals between the two wake ups follow an exponential distribution. Besides, the second awake stage generally reaches its peak at a higher velocity and higher volume of the first spike usually lead to much higher popularity of the second spike. Based on these findings, we model the two sleeping beauties based on the classical Bass diffusion model(~\cite{bass1969,Norton1987}) and make well descriptions for the meme diffusion. We disclose a common mechanism beyond meme diffusion in terms of two sleeping beauties and our results can help reveal the fundamental rule that drives the popularity dynamics of memes in different media.

\section{Materials and Methods}
\label{sec:mm}

\subsection{Datasets}

The experiment is conducted on three datasets. The first dataset is metadata for the complete set of all PubMed records from 1809 to 2012 (with part of 2013 available as well), including title, authors, and year of publication~\cite{Light2014data,LaRowe2008data}. We segment the title into $n$-grams ($n$=1, 2, 3) and by finding $n$-grams appearing at least 50 times, we finally get 86,784 uni-grams, 190,992 bi-grams, and 47,646 tri-grams. It should be noted that even the frequency of these $n$-grams in paper title is more than 50, there still exit a large proportion of meaningless phrases. From the total 325,422 $n$-grams, we obtained 741 memes with two remarkable sleeping beauties in the diffusion.

The other two datasets are obtained from Wikipedia and Google Trends respectively, which are publicly available from ~\cite{Yoshida2015}. The search frequency of an entry in Wikipedia or a query in Google generally reflects the collective attention on a particular subject in the real world~\cite{Kristoufek2013}. So both of Wikipedia page views and Google Trends are valuable resources for information diffusion study. In the two datasets, the search queries set contains 3,231 keywords about Cartoon, 7,251 keywords about comic, 10,000 keywords about movie and 10,000 celebrity names. These queries were accessed from 2008 to 2014 and their daily, weekly and monthly search volumes in Google Trends and page views in Wikipedia were all collected by ~\cite{Yoshida2015}. It should be noted that the data in Google Trends is a little different from that Wikipedia. Google trends are expressed as the percentage integers of a maximum value for a particular search interval, and the trends of low frequency keywords consist mostly of zeros or are not even provided by Google Trends.

\begin{table}[ht]
\centering
\caption{
Statistics of the datasets}
\label{table1}
\begin{tabular}{lllllll}
\hline
\multicolumn{1}{l}{Time Granularity} &\multicolumn{2}{l}{Google Trends} & \multicolumn{2}{l}{Wikipedia} & \multicolumn{2}{l}{Paper Titles}\\ \hline
Day & &  & 29072 & {\bf 1859} &  & \\ \hline
Week & 13195 & {\bf 1402} & 29072 & {\bf 3815} &  & \\ \hline
Month & 22535 & {\bf 1641} & 29072 & {\bf 1946} &  & \\ \hline
Year &  &  &  &  & 325422 & {\bf 741}\\ \hline
\end{tabular}
\begin{flushleft}For each dataset, the left column denotes the number of all memes and the right column in bold denotes the number of memes with two remarkable sleeping beauties.
\end{flushleft}
\end{table}

In total, we get a large number of memes with two sleeping beauties from the three datasets and the detailed statistics are summarized in Table \ref{table1}. The fraction of memes with two sleeping beauties ranges from 5\% to 10\% in Wikipedia page views and Google search queries, which agrees well with the foundings in~\cite{science2015Wang}. Note that because of noisy grams in paper titles, the ratio of memes with two sleeping beauties in the dataset of $n$-grams is relatively low. 

\subsection{Identification of the two sleeping beauties}
Inspired by ~\cite{Palshikar2009} and ~\cite{Ke2015}, we develop methods to detect and measure the phenomenon of two sleeping beauties in memes diffusion. In addition, the measurement solution is parameter independent and can be easily extended to different domains.

The first step of identifying sleeping beauties is to locate candidate peaks in the meme's popularity dynamics by successfully filtering out the noisy cascades with trivial popularity peaks. ~\cite{Palshikar2009} proposed a simple but effective algorithm for peak detection from a number of noisy cascades. By using this method with $k = 5$ and $h = 0.5$ (the parameters selection is discussed by ~\cite{Palshikar2009} and the parameters are universal for different kinds of time series), we first locate the highest peak ($P_2$) from the popularity dynamics and then target the other candidate meaningful peaks that happens before $P_2$. If there exists at least two peaks, the meme will be labeled as candidate sleeping beauties. Then, we filter sleeping beauties according to the beauty coefficient of different peaks for each candidate meme.

\begin{figure}[h]
\begin{center}
\noindent
\includegraphics[width=0.8\linewidth]{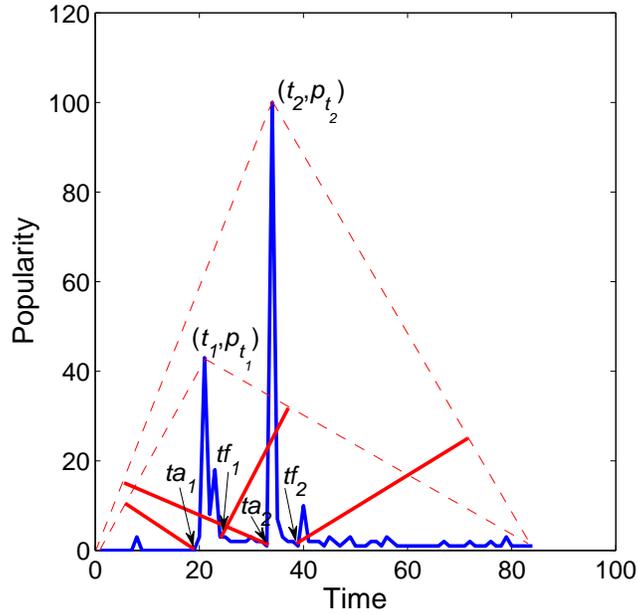}
\caption{Illustration of the identification of two awakening times $ta_{1}$ and $ta_{2}$, and two falling asleep times $tf_{1}$ and $tf_{2}$. The blue curve represents the popularity dynamics of a meme at age $t$ (i.e., $t$ represents the time since its appearance). Dotted lines represent the auxiliary lines. The red thick lines represent the distance from the point $(t, S(t))$ to the auxiliary line.}
 \label{fig2}
\end{center}
\end{figure}

According to three dimensions of describing sleeping beauties, i.e., the length of sleep, the depth of sleep, and the awake intensity presented by ~\cite{Van2004}, ~\cite{Ke2015} proposed an index named Beauty coefficient ($B$) and then introduced a parameter-free framework to measure the sleep beauty. Inspired by their approach, as illustrated in Fig.\ref{fig2}, we further establish a method to measure the phenomenon of two sleeping beauties. Given a meme, we define $S(t)$ as the popularity at time $t$ after its generation, i.e., $t$ stands for the age of the meme and the diffusion starts at $t_0 = 0$ and ends at time $T$. 

Assume that the meme receives its maximum popularity at time $t_2$, then the straight line that connects points ($t_0$, $S(t_0)$) and ($t_2$, $S(t_2)$) in the time-popularity plane, denoted as $la_{t}$, can be depicted as
\begin{equation}\label{eq:scheme1} 
la_t=\frac{S(t_2)-S(t_0)}{t_2-t_0}(t-t_0)+S(t_0).
\end{equation}

 The awakening time for peak $P_2$, denoted as $ta_2$, can be defined as the time $t$ at which the distance ($da_t$) between point ($t$, $S(t)$) and the reference line $la_{t}$ reaches the maximum. Specifically, we have
\begin{equation}\label{eq:scheme2} 
ta_{2}=\arg\left\{\max_{t<t_2}da_t\right\},
\end{equation}
where
\begin{equation}\label{eq:scheme3} 
da_t=\frac{\left|(S(t_2)-S(t_0))(t-t_0)+(t_2-t_0)(S(t_0)-S(t))\right|}{\sqrt{(S(t_2)-S(t_0))^2+(t_2-t_0)^2}}.
\end{equation}

Similarly, the straight line that connects the points ($T$, $S(T)$) and ($t_2$ , $S(t_2)$) in the time-popularity plane, denoted as $lf_{t}$, can be formulated as
\begin{equation}\label{eq:scheme4} 
lf_t=\frac{S(t_2)-S(T)}{t_2-T}(t-T)+S(T).
\end{equation}

Then the time of falling asleep, denoted as $tf_2$, can be taken as the time $t$ at which the distance $df_t$ between the point ($t$, $S(t)$) and the reference line $lf_{t}$ reaches the maximum. Accordingly, we have
\begin{equation}\label{eq:scheme5} 
tf_{2}=\arg\left\{\max_{t>t_2}df_t\right\},
\end{equation}
where 
\begin{equation}\label{eq:scheme6} 
df_t=\frac{\left|(S(t_2)-S(T))(t-T)+(t_2-T)(S(T)-S(t))\right|}{\sqrt{(S(t_2)-S(T))^2+(t_2-T)^2}}.
\end{equation}

Again, considering the straight line that connects points (0, $S(t_0)$) and ($t_1$ , $S(t_1)$) and the straight line that connects points ($T$, $S(T)$) and ($t_1$ , $S(t_1)$) in the time-popularity plane, the awakening time $ta_1$ and the falling asleep time $tf_1$ can be similarly located.

Then, we get eight time stamps, i.e., $t_0$, $ta_1$, $t_1$, $tf_1$, $ta_2$, $t_2$, $tf_2$ and $T$, and these time stamps shall be in ascending order according to their value, otherwise the corresponding candidate will be excluded. Based on these time stamps, we can compute the Beauty coefficient $B_i(i = 1,2)$ for the two sleeping beauties of filtered candidate memes. 
\begin{equation}\label{eq:scheme7} 
B_1=\frac{1}{ta_1-t_0}\sum_{t=t_0}^{ta_1}\frac{\frac{S(t_1)-S(t_0)}{(t_1-t_0)}(t-t_0)+S(t_0)-S(t)}{max\{1,S(t)\}}.
\end{equation}
\begin{equation}\label{eq:scheme8} 
B_2=\frac{1}{ta_2-tf_1}\sum_{t=tf_1}^{ta_2}\frac{\frac{S(t_2)-S(tf_1)}{(t_2-tf_1)}(t-tf_1)+S(tf_1)-S(t)}{max\{1,S(t)\}}.
\end{equation}

By definition, memes with popularity growing linearly with time ($lt = S(t)$ ) have $B$ = 0. and $B$ is non-positive for papers whose citation trajectory $la_t$ is a
concave function of time. When both of $B_1$ and $B_2$ are greater than $\alpha S(t_1)$ and $\alpha S(t_2)$ ($\alpha = 1/3$, $\alpha$ is also universal for memes in different media) respectively, the candidate meme will be identified as sleeping beauty.

\section{Results}
\label{sec:res}

The probe on different datasets demonstrates that the two sleeping beauties are pervasively existing in different media and can be convincing proxy to explore the common mechanism beyond diffusion of different memes. Meanwhile, it is also demonstrated that the time intervals between two wake ups follow an exponential distribution, the propagation velocity in the second sleeping beauty will be much higher than the first one and the total popularity during the second wake up is positively related to the first wake up. These key characteristics of the sleeping beauties can be employed to establish a mathematical model to replicate the popularity dynamics of memes from different media.

\subsection{Exponential intervals between wake ups}

For each meme with two sleeping beauties in diffusion, we can obtain the first awakening time $ta_1$, the time of the first peak popularity $t_1$, the first falling asleep time $tf_1$, the second awakening time $ta_2$, the time of the second peak popularity $t_2$ and the second falling asleep time $tf_2$. We define the time interval between two wake ups as $ta_2-tf_1$, which reflect the length of the second sleeping. Assuming the first sleeping beauty is observed, then the time interval can be employed to predict the second awaking time, i.e., when the meme will start to experience a new spike of popularity. 

\begin{figure}[ht]
\centering
\includegraphics[width=\linewidth]{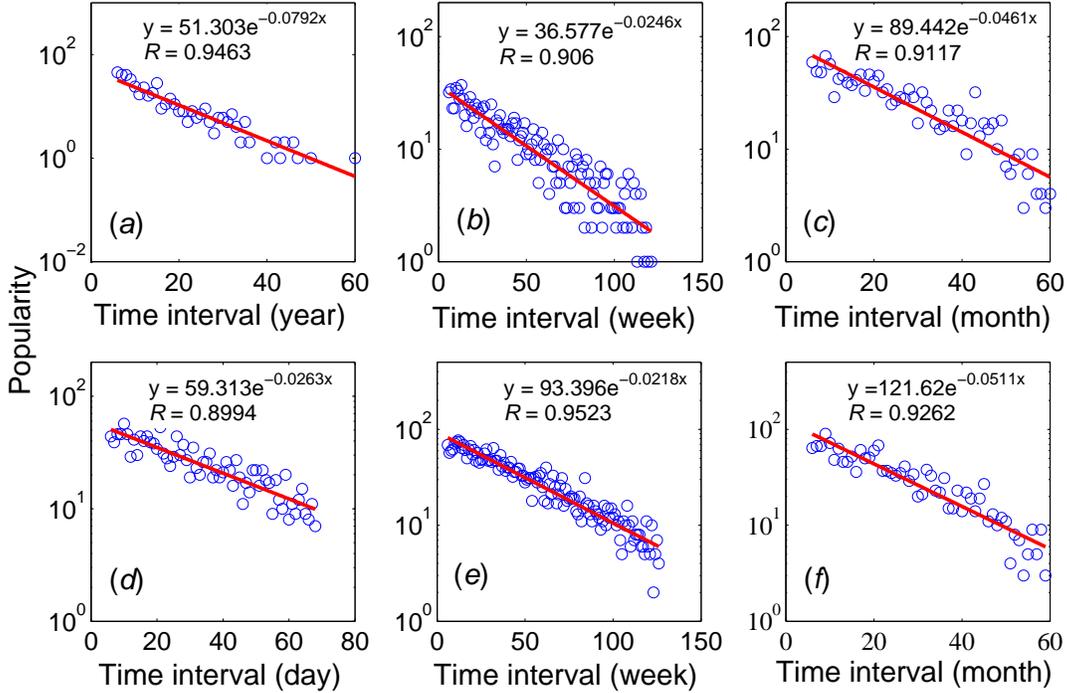}
\caption{Distribution of time intervals between wake ups. $R$ denotes Pearson product-moment correlation coefficient and higher values stand for better fittings. (\emph{a}) $N$-grams of publication titles. (\emph{b} and \emph{c}) Search queries of Google Trends. Here the popularity is defined as the normalized search frequency in Google. (\emph{d}, \emph{e} and \emph{f}) Wikipedia page views.}
\label{fig3}
\end{figure}

We measure the time intervals between two wake ups for different memes in our datasets and surprisingly find that they follow stable exponential distributions in different media and the coefficients are very close to each other. As can be seen in Fig.~\ref{fig3}, $\lambda$, the exponential coefficient, is respectively 0.0792, 0.0246, 0.0461, 0.0263, 0.0218 and 0.0511 for different datasets with different time granularities. Considering the fact that exponential distribution has the key property of being memoryless, the above finding suggests that only temporal patterns may not be enough to predict the second awakening. Meanwhile, note that $\lambda$ in different datasets mainly locates in the narrow range of $[0.02,0.08]$, which indicating that even for different media, the distribution is almost the same and it further supports our hypothesis that there exists a common and media-independent mechanism beyond the meme diffusion. 

\subsection{The comparison between two wake ups}
\begin{figure}[ht]
\noindent
\includegraphics[width=\linewidth]{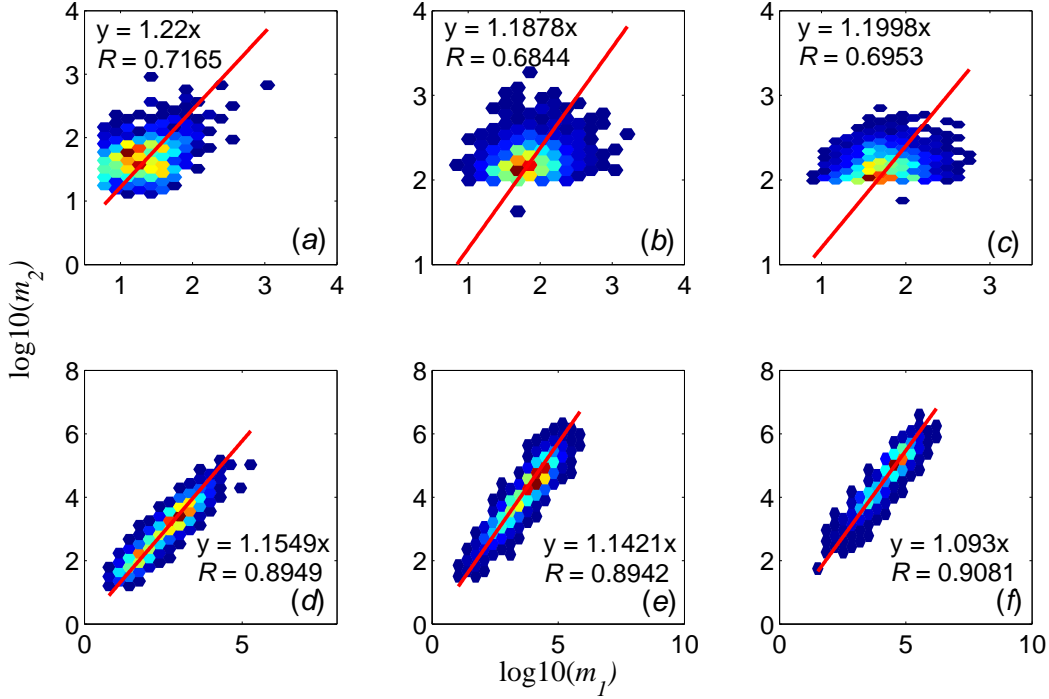}
\caption{ Correlation between the total popularity of two wake ups. $m_1$ is the total popularity during the first wake up and $m_2$ is the total popularity in the second wake up. $R$ denotes Pearson product-moment correlation coefficient. Brigher hexbins means higher frequency. (\emph{a}) $N$-grams in publication titles. (\emph{b} and \emph{c}) Search queries from Google Trends with time granularities of week and month. Popularity denotes the normalized search frequency in Google. (\emph{d}, \emph{e} and \emph{f}) Wikipedia page views with multiple time granularities of day, week and month.}
\label{fig4}
\end{figure}


\begin{figure}[ht]
\noindent
\includegraphics[width=\linewidth]{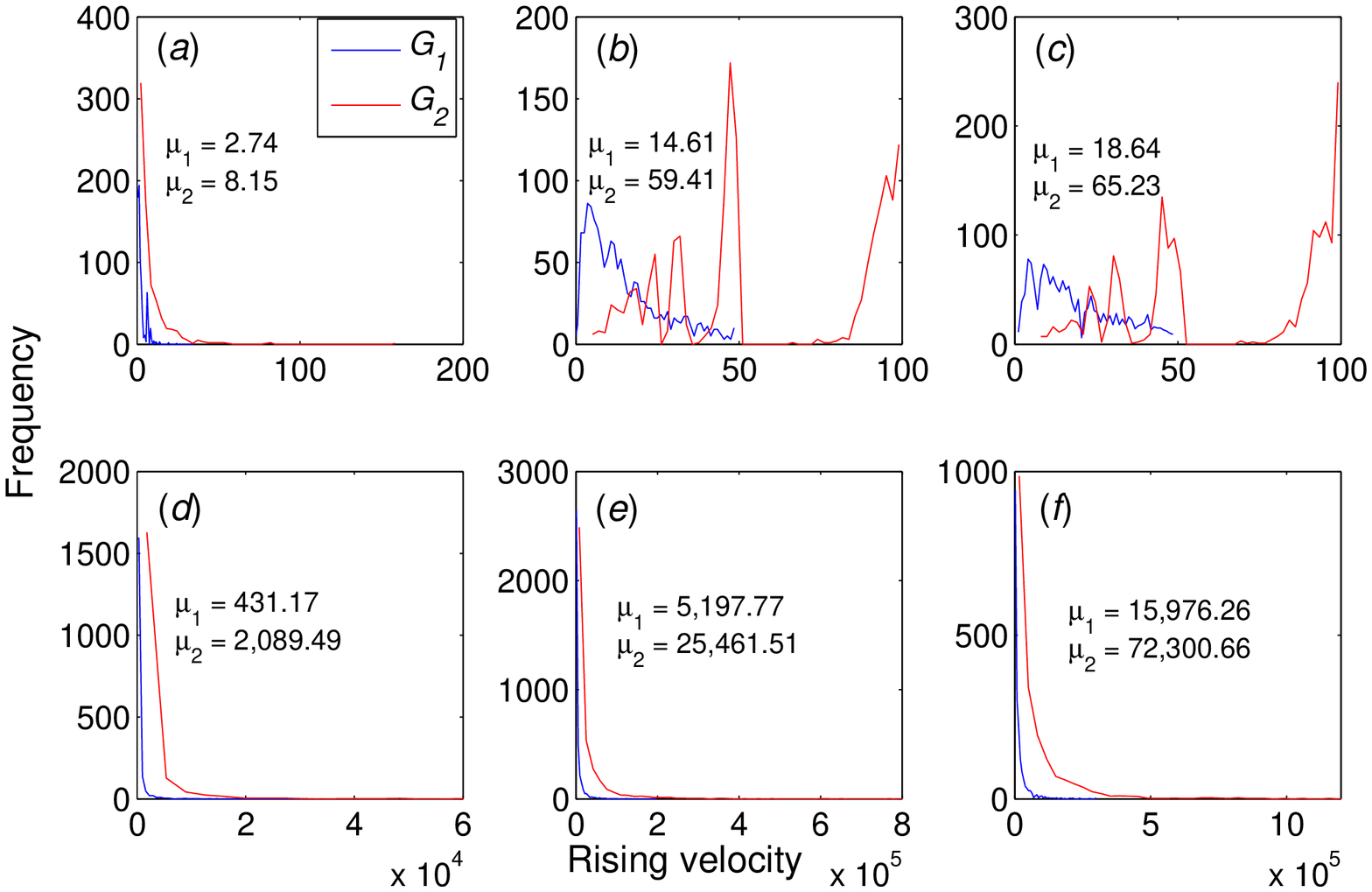}
\caption{ Comparison of rising velocity in wake up. $\mu_1$ and $\mu_2$ denote the average rising velocity in $G_1$ and $G_2$. (\emph{a}) $N$-grams of publication titles. (\emph{b} and \emph{c}) Search queries from Google Trends with time granularities of week and month. (\emph{d}, \emph{e} and \emph{f}) Wikipedia page views with multiple time granularities of day, week and month.}
\label{fig5}
\end{figure}


During a wake up, the meme obtains collective attention and shows popularity spikes. As can be seen in Fig.~\ref{fig4}, the comparison of the total popularity, i.e., the total occurrences in paper titles or the total search volume in Google and Wikipedia page views shows that more attention in the first wake up will lead to even more popularity in the second wake up. However, it should be noted that for the dataset of Google Trends, there only exists weak correlation between $m1$ and $m2$, it perhaps because that the popularity value in Google Trends has been normalized or transformed and the popularity dynamics of low frequency keywords are not even provided. In detail, we can formalized the relation between the total popularity between two wake ups, i.e., $m_2=m_1^\alpha$, where $\alpha$ locates in the narrow range of [1.093,1.22]. Meanwhile, another metric to reflect the formation of the popularity spike is the rising velocity, which can be directly defined as $v_i=(S(t_i)-S(ta_i))/(t_i-ta_i)$, where $S(t)$ denotes the popularity of the meme at $t$. As can be seen in Fig.~\ref{fig5}, the average rising velocity in the second wake up is 3-5 times faster than the one in the first wake up. From the comparison of two wake ups, we can find that the early rising velocity is a predictive feature of cascade size, which is consistent with the finding in~\cite{cheng2014can}.

\subsection{Modeling the popularity dynamics of sleeping beauties}

With the aim of modeling popularity dynamics of different memes in different media, we have to neglect many detailed and domain-dependent factors like community structure, homogeneity or competition and focus on establishing a general framework only based on the statistics from the two sleeping beauties. Therefore, we upgrade the classical diffusion model named Bass model, which was developed by ~~\cite{bass1969} and ~\cite{Norton1987}, to model the memes' popularity dynamics.
\begin{figure}[ht]
\noindent
\includegraphics[width=\linewidth]{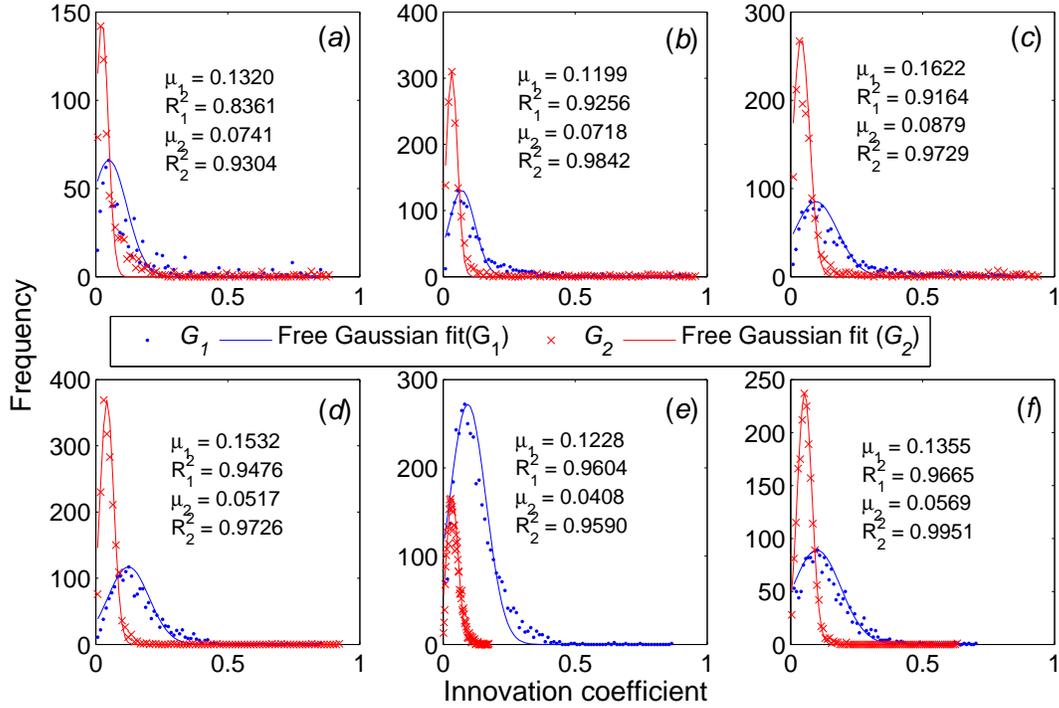}
\caption{ Free Gaussian fit of $p_i$ from different datasets. The density function of Free Gaussian is $y=a_i*exp(-(x-\mu_i)^2/(2*\sigma_i^2))$, $\sigma_i$ and $\mu_i$ are standard deviation and mean of innovation coefficient in $G_i$. $R^2_i$ denotes the coefficient of determination between free Gaussian fit and $G_i$. (\emph{a}) $N$-grams of publication titles. (\emph{b} and \emph{c}) Search queries of Google Trends with time granularities of week and month. (\emph{d}, \emph{e} and \emph{f}) Wikipedia page views with time granularities of day, week and month.}
\label{fig6}
\end{figure}

Bass model originally depicts the diffusion of innovation and imitation. Specifically, innovators create innovation and the other individuals in the social system might adopt the innovation at different time. Considering the pervasive existing of two sleeping beauties in the lifetime of memes, here we correspondingly separate the entire diffusion into two generations $G_i(i=1,2)$. Let $S_i(t)$ be the popularity of a meme at time $t$ during its $i_{st}$ generation, which can be obtained from 
\begin{equation}\label{eq:scheme9} 
S_1(t)=m_1F_1(t)-m_1F_1(t)F_2(t-ta_2)=m_1F_1(t)(1-F_2(t-ta_2)),
\end{equation}
and 
\begin{equation}\label{eq:scheme10} 
S_2(t)=m_2F_2(t-ta_2)+m_1F_1(t)F_2(t-ta_2)=F_2(t-ta_2)(m_2+m_1F_1(t)),
\end{equation}
$m_i$ represents the diffusion potential for the $i_{st}$ diffusion of the meme, while $F_i(t)$ is the diffusion rate of the $i_{st}$ diffusion at time $t$ and can be evaluated by 
\begin{equation}\label{eq:scheme11} 
F_i(t)=\left\{
\begin{aligned}
&0, && t<0 \\
&\frac{1-e^{-(p_i+q_i)t}}{(q_i/p_i)e^{-(p_i+q_i)t}+1}, && t\geq 0 \\
\end{aligned}
\right.
\end{equation}
in which $p_i$ is the innovation coefficient and $q_i$ denotes the imitation coefficient. For the case of two sleeping beauties, the innovation coefficient can be calculated through
\begin{equation}\label{eq:scheme12} 
p_i = \frac{S_i(ta_i)+S_i(ta_i+1)}{2m_i}.
\end{equation}
Note that the approach of sleeping beauty identification (see Methods) tends to pick the time with lowest and nearest popularity before $t_1$ as the awakening time, indicating that the popularity at $ta_i$ is generally close or equal to zero. In order to avoid the problem, it is reasonable to consider the popularity at the next point cooperatively, i.e., $ta_i+1$, to estimate the innovation coefficient. And the imitation coefficient $q_i$ can be calculated from
\begin{equation}\label{eq:scheme13} 
t_i-ta_i = \frac{1}{p_i+q_i}ln(q_i/p_i).
\end{equation}

\begin{figure}[ht]
\noindent
\includegraphics[width=\linewidth]{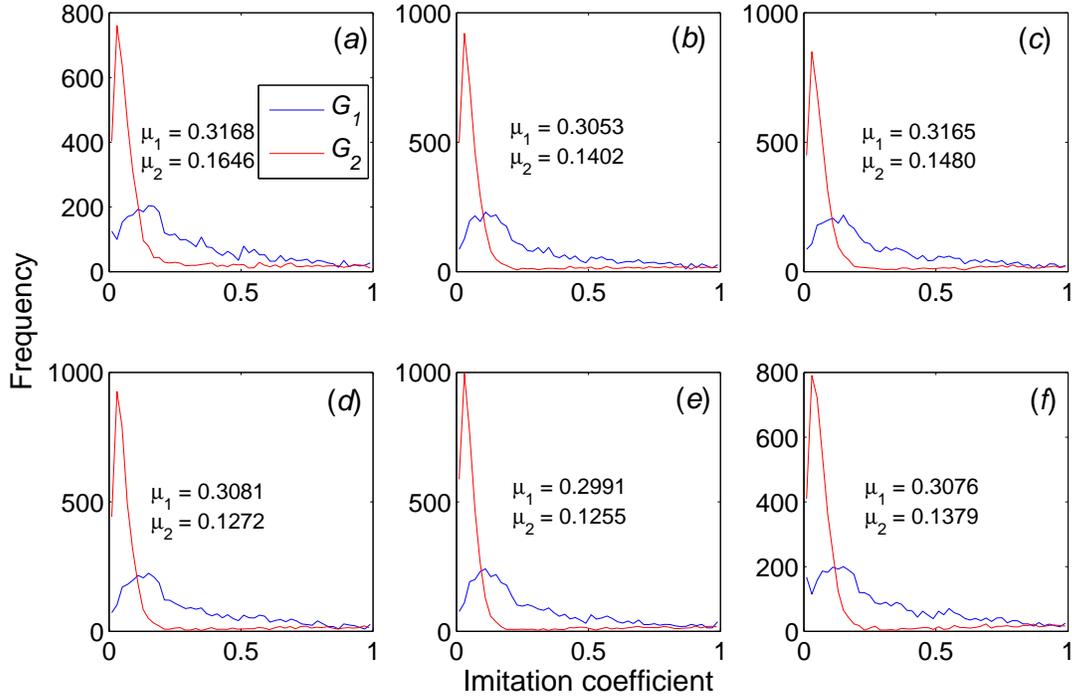}
\caption{ Distribution of $q_i$ from different datasets.
$\mu_i$ is the average imitation coefficient in $G_i$. (\emph{a}) $N$-grams of publication titles. (\emph{b} and \emph{c}) Search queries of Google Trends with time granularities of week and month. (\emph{d}, \emph{e} and \emph{f}) Wikipedia page views with time granularities of day, week and month.}
\label{fig7}
\end{figure}

In meme diffusion, $p_i$ can model the influence brought by external factors (e.g. stimuli) while $q_i$ in fact represents the internal factors (e.g. social networks) that might also shape the popularity dynamics. We fit the distribution of innovation coefficient according to Log-normal function and free Gaussian function respectively. And the coefficient of determination suggest that the overall fitting goodness of free Gaussian distribution is better than Log-normal. With respect to $p_i$, as can bee in Fig.~\ref{fig6}, it is surprising to find that all the memes' $p_i$ follows a Gaussian-like distribution and the first beauty possesses a higher averaged value than the second.  The Gaussian-like distribution, especially the mean locating in a narrow range, further testify our assumption that different memes from different media are driven by a common mechanism to attract collective attention. And higher $p_i$ for the first beauty also suggests that for each meme, the external factors functions more significantly in the first beauty than in the second. Regarding to imitation coefficient $q_i$, as seen in Fig.~\ref{fig7}, the average value in the first beauty is about two times higher than the second one. However, higher imitation coefficient cannot guarantee broader diffusion, because the imitation pressure at $t_n$ from the internal network can be quantified as $(q_i/m_i)\sum_{t=1}^{t_n}{S_i(t)}$ while $\sum_{t=1}^{t_n}{S_i(t)}$ is greatly determined by the rising velocity in the awakening period and as shown in Fig.~\ref{fig5}, the average rising velocity in the second beauty is about five times higher than the first wake up.

In the conventional parameter estimation, Bass model infers the innovation coefficient, imitation coefficient and total diffusion amount from the first more than three observations of the diffusion. However, because many memes in different media usually reach their peak popularity in no more than three time units, the previous method is not applicable here and we have to present a new estimation approach. We set the first awakening time as the start timing for each meme and $ta_1$, $t_1$, $tf_1$, $ta_2$, $t_2$ and $tf_2$ can be obtained from each sleeping beauty's popularity history. Gaussian-like distribution of $p_i$ implies that the average value from historical observations can be the estimation for the innovation coefficient and accordingly the imitation coefficient can be obtained through Eq.~\ref{eq:scheme13}. Then the total popularity $m_i$ for $G_i$ can be estimated by $\sum_{t=ta_i}^{tf_i}{S_i(t)}$. 

\begin{figure}[ht]
\noindent
\includegraphics[width=\linewidth]{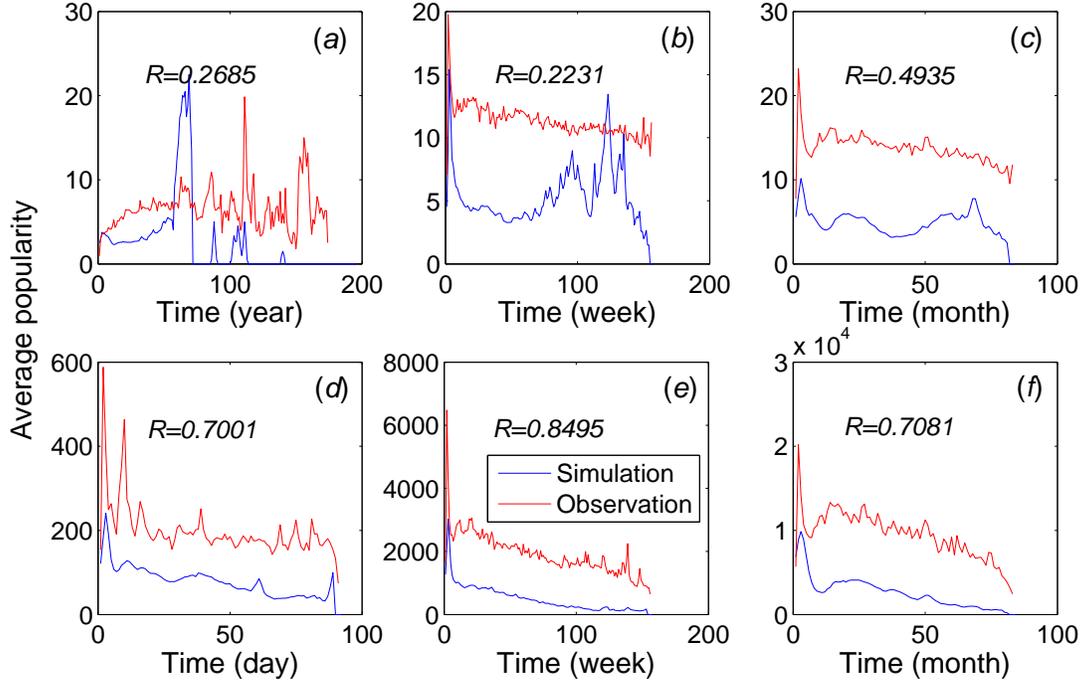}
\caption{ Evaluation of simulation with average innovation coefficient. The innovation coefficient is set to the mean of all memes in each dataset. $R$ denotes Pearson correlation coefficient between the the observed curve and the simulation curve. (\emph{a}) $N$-grams of publications titles. (\emph{b} and \emph{c}) Search queries of Google Trends with time granularities of week and month. Popularity denotes the normalized search frequency in Google. (\emph{d}, \emph{e} and \emph{f}) Wikipedia page views with multiple time granularities of day, week and month.}
\label{fig8}
\end{figure}

To facilitate the model evaluation, the observed trend of the popularity is averaged over different memes and the average popularity at $t$ is defined as $\left\langle{S_i}(t)\right\rangle=1/n\sum_{j=1}^{m}S_{ij}(t)$, where $n$ stands for the number of memes whose $S_i(t)\neq0$ and $m$ is the total number of sleeping beauties. Then its similarity (e.g. Pearson correlation coefficient) to the averaged simulation dynamic will be used to vividly demonstrate the model performance. 

\begin{figure}[ht]
\noindent
\includegraphics[width=\linewidth]{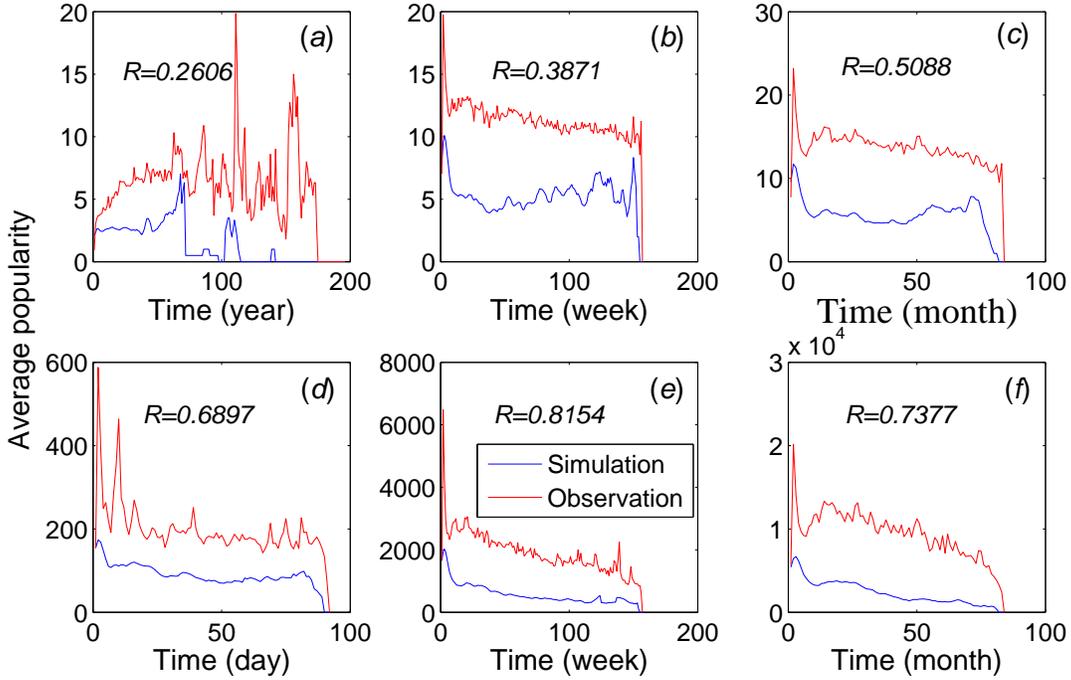}
\caption{ Evaluation of the simulation with observed innovation coefficient. The innovation coefficient is set to the observed value of each meme. (\emph{a}) $N$-grams of publications titles. (\emph{b} and \emph{c}) Search queries of Google Trends with time granularities of week and month. Popularity denotes the normalized search frequency in Google. (\emph{d}, \emph{e} and \emph{f}) Wikipedia page views with multiple time granularities of day, week and month.}
 \label{fig9}
\end{figure}

As can be seen in Fig.~\ref{fig8}, we obtain well simulation for the locations of different beauties from different datasets and the Pearson correlation coefficient between the simulated and observed popularity dynamics in monthly Google trends and Wikipedia page views are particularly high. The result in Fig.~\ref{fig9} shows the simulation by setting each sleeping beauty's innovation coefficient to its observed value instead of the average (the setting of imitation coefficient is same), the minor improvement further suggests the validity of using the mean to estimate the innovation coefficient for different memes, implying that our approach is robust to memes, domains and time granularities.

\begin{table}[ht]
\centering
\caption{
The accuracy of simulation for sleeping beauties' second popularity peak timing}
\label{table2}
\begin{tabular}{lllll}
\hline
{$p@k$}   & {$p@0$}   & {$p@1$}   & {$p@2$}    & {$p@3$} \\ \hline
$\#$memes & 3814 & 8661 & 10611 & 11010 \\ \hline
Precision & 33.44\% & 75.95\% &  93.05\% & 96.55\% \\ \hline
\end{tabular}
\begin{flushleft} $k$ is defined as the absolute value of the difference between the simulated peak timing and the observed peak timing. $\#memes$ represent the number of memes as the difference is smaller than $k$. $p@k$ is defined as the precision of the simulation as the difference is smaller than $k$ and higher values indicate better simulations.
\end{flushleft}
\end{table}

\begin{figure}[ht]
\begin{center}
\noindent
\includegraphics[width=0.8\textwidth]{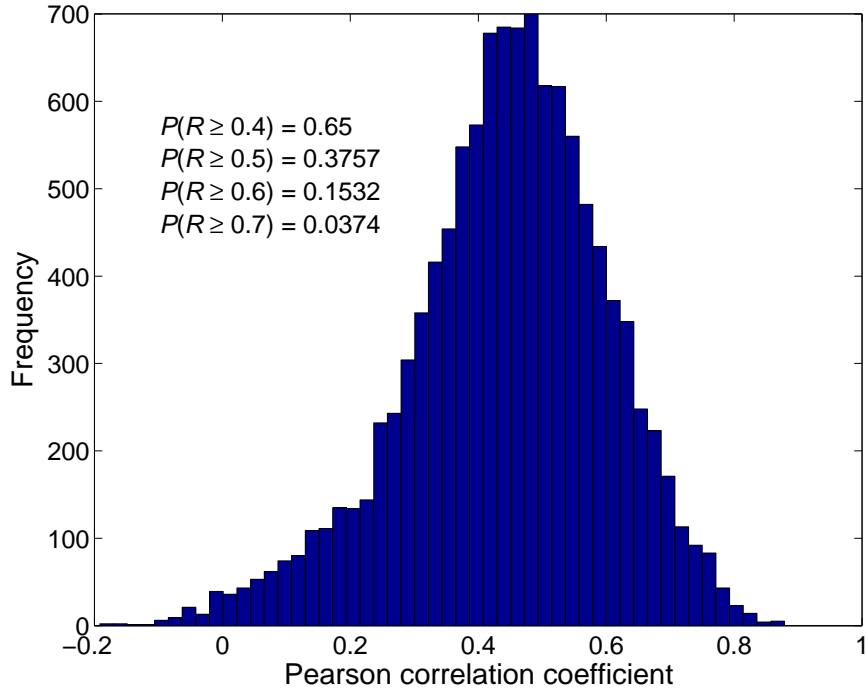}
\caption{ The Pearson correlation coefficient of each meme's fit. The innovation coefficient is set to the mean of all memes in each dataset.}
 \label{fig10}
 \end{center}
\end{figure}

To evaluate the fitting effect of each meme, we also use Pearson correlation coefficient measure the simulated dynamic time series and the observed data of all the 11400 sleeping beauties in these datasets. A high value of correlation means the simulated and the observed data are strongly correlated. As shown in Fig.~\ref{fig10}, The expectation of correlation is about 0.5 and there exists a relative 65\% of simulated curves with correlation greater than 0.4. What's more, we also compute the precision of simulation for each meme's second peak time. As can be seen in Table.~\ref{table2}, the accuracy of the unbiased simulation for the second peak time is 33.44\% and the accuracy of simulation with deviation less than two time units reach 75.95\%. The results suggest the validity of our framework, especially for the precision of over 96\% as the deviation less than three time units, which is in fact generally acceptable in the realistic scenarios.

Note that we may not get the precise estimation of the absolute value of the popularity at certain time, but we aim to give a promising description of where two sleeping beautifies will locate. What's more, the volume difference between the simulation and the observation is partly caused by that the popularity in sleeping period is set to zero. The above results indicate that based on the statistics of two sleeping beauties in different memes, we can use the upgraded Bass model to replicate the popularity trend for memes on different media, especially the location of two beauties, which can help identify key points in the diffusion.

\section{Discussion}
\label{sec:dis}

By providing detailed tracks of diffusion, Internet indeed offers us an unprecedented opportunity to deeply understand the dynamics of information. However, the previous study tends to focus on different factors influencing the meme propagation, for examples, the intrinsic infectiousness, system network structure and time-spatial factors. These specific features or factors are usually dependent on system. Using nonparametric method model which implicitly accounts for all these factors might be a better way to investigate the propagation of different information. Motivated by this, we try to disclose the common mechanism from the perspective of sleeping beauties, which is formerly studied in citation history of scientific publications. 

Surprisingly, we find the sleeping beauty, especially the phenomenon of two consecutive sleeping beauties, is pervasively existing in diffusion of different information from different media. We systematically explore the basic features of two beauties and establish a Bass-model-based approach to replicate the popularity dynamics. The promising result of our model solidly demonstrates that the mechanism behind different memes exists and its function in driving the formation of popularity dynamics can be reflected by the two sleeping beauties we reveal.

Even more inspiring, our investigations also show that for memes with two sleeping beauties, the higher volume of the first spike will lead to even much higher popularity of the second spike with great odds, i.e., it leads to the outstanding peak in the lifetime of diffusion. Recall that the outstanding peak is conventionally thought to be caused by the tipping point of the meme or the prince of the sleeping beauty, hence our approach can help locate the timing of the tipping point in specific media, which is intuitively the awakening time of the second beauty. So from this view, our findings and solutions can be important to practical applications like marketing business and scientific impact prediction.

However, our work has inevitable limitations. The phenomenon of two sleeping beauties is a typical form of multiple sleeping beauties. There exist many memes possessing more than twice sleeping beauties. For simplicity, we only investigate the two remarkable beauties of memes from different datasets. However, further explorations in the phenomenon of multi sleeping beauties would be promising direction in the future work.

In summary, the mechanism beyond the diffusion dynamics of memes might be complicated and hard to be tracked, however, we argue that the phenomenon of two sleeping beauties could provide a new and insightful view to understand it. And in the future work, we would like to concentrate on the memes with more than twice sleeping beauties and develop more detailed models to capture the fine-granular dynamics of memes and provide the prediction both in volume and timing simutaneously. 
 
\section*{Acknowledgment}
\label{sec:ack}
This work was supported by the National Natural Science Foundation of China (Grant Nos. 71501005 and 71531001) and the fund of the State Key Lab of Software Development Environment (Grant Nos. SKLSDE-2015ZX-05 and SKLSDE-2015ZX-28). We also thank Ms. Xiaoqian Hu for her valuable suggestions.

\bibliographystyle{plain}
\bibliography{refs}

\section*{Supporting Information}
{\bf S1 Datasets and code.} The datasets and code employed in the research can be available at: \url{https://figshare.com/articles/Meme_popularity_and_diffusion/3187159/1}.

\end{document}